\documentclass[12pt]{article}
\usepackage{amsfonts}
\usepackage{amsmath}
\usepackage{amssymb}
\usepackage{axodraw}
\usepackage{mathptmx}
\usepackage[dvips]{color}
\usepackage{epsfig}
\usepackage{graphicx}
\newcommand{\calL}{{\cal L}}
\newcommand{\calC}{{\cal C}}
\newcommand{\calG}{{\cal G}}
\newcommand{\calM}{{\cal M}}
\newcommand{\calU}{{\cal U}}
\newcommand{\calW}{{\cal W}}
\newcommand{\calX}{{\cal X}}
\newcommand{\calY}{{\cal Y}}
\newcommand{\calZ}{{\cal Z}}

\newcommand{\dslash}{/\!\!\!\!\!\partial}
\newcommand{\Dslash}{/\!\!\!\!\!D}

\newcommand{\Li}{\textrm{Li}}

\newcommand{\eV}{{\rm eV}}
\newcommand{\GeV}{{\rm GeV}}

\begin{document}
\baselineskip=16pt

\pagenumbering{arabic}

\vspace{1.0cm}

\begin{center}
{\Large\sf Radiative neutrino mass in type III seesaw model}
\\[10pt]
\vspace{.5 cm}

{Yi Liao$^{a,b,c}$\footnote{liaoy@nankai.edu.cn}, Ji-Yuan Liu$^a$,
Guo-Zhu Ning$^a$}

{$^a$ Department of Physics, Nankai University, Tianjin 300071,
China\\
$^b$ Center for High Energy Physics, Peking University, Beijing
100871, China\\
$^c$ Kavli Institute for Theoretical Physics China, CAS, Beijing
100190, China}

\vspace{2.0ex}

{\bf Abstract}

\end{center}

The simplest type III seesaw model as originally proposed introduces
one lepton triplet. It thus contains four active neutrinos, two
massive and two massless at tree level. We determine the radiative
masses that the latter receive first at two loops. The masses are
generally so tiny that they are definitely excluded by the
oscillation data, if the heavy leptons are not very heavy, say,
within the reach of LHC. To accommodate the data on masses, the
seesaw scale must be as large as the scale of grand unification.
This indicates that the most economical type III model would entail
no new physics at low energies beyond the tiny neutrino masses.

\begin{flushleft}
PACS: 14.60.Pq, 14.60.St, 12.15.Lk, 14.60.Hi

Keywords: radiative neutrino mass, seesaw model, lepton triplet

\end{flushleft}

\newpage

\section{Introduction}
\label{sec:intro}

The standard model (SM) of electroweak interactions when viewed as
an effective field theory at low energies, has a unique dimension
five operator that can generate Majorana neutrino masses
\cite{Weinberg:1979sa}. And the operator has only three possible
realizations at tree level \cite{Ma:1998dn}. These correspond to the
celebrated three types of seesaw models
\cite{type1,type2,Foot:1988aq}. While the type I model introduces
sterile neutrinos as the minimal option to operate the seesaw, the
other two prescribe particles that participate electroweak
interactions. If the seesaw scale is not too high, richer phenomena
are expected in the last two types of models. There have been
extensive investigations on the type I and II seesaw models, but the
interest in type III has been catalyzed recently by the advent of
the LHC at CERN \cite{Chen:2009vx}, where the assumed triplet
leptons could be directly produced through gauge interactions if
they are not too heavy
\cite{Bajc:2007zf,Franceschini:2008pz,delAguila:2008hw,Blanchet:2008zg}.
Various other phenomenological aspects of the model have also been
explored, including possible modifications to leptogenesis
\cite{Hambye:2003rt,Fischler:2008xm,Strumia:2008cf,Blanchet:2008ga,
Blanchet:2008zg}, low energy effects in lepton flavor changing
processes \cite{Abada:2007ux} and anomalous magnetic moments of
charged leptons \cite{Chao:2008iw,Biggio:2008in}, renormalization
group running of neutrino parameters \cite{Chakrabortty:2008zh}, and
the potential role as dark matter \cite{Ma:2008cu}, to mention a
few.

For a seesaw model like type III to be relevant at relatively low
energies, it must be capable of incorporating the data from
oscillation experiments and other constraints with a not too high
seesaw scale. We are thus motivated to start with the simplest type
III seesaw as was originally proposed \cite{Foot:1988aq}. It extends
SM by one triplet of leptons, resulting in two massive and two
massless neutrinos at tree level, plus a pair of heavy charged
leptons. It also serves as an approximation to more general
structures that contain additional sequentially heavier triplets of
leptons. The massless neutrinos not being protected by any symmetry
should receive radiative masses, which will be determined in this
work. It would be interesting to ask whether it is possible in this
minimal model to get a radiative mass at a desired level with a
seesaw scale accessible at LHC.

The idea of generating a one-loop radiative mass for neutrinos was
originally suggested in Ref \cite{Zee:1980ai}, and extended to two
loops in \cite{Zee:1985id,Babu:1988ki}. It offers a nice way to
induce hierarchical and tiny neutrino masses. There is a vast
literature that extends the idea in various aspects (see as
examples,
\cite{Ma:1997nq,Grimus:1999wm,Chang:1999hga,Lavoura:2000kg,
Kitabayashi:2000nf}) and calculates radiative masses in different
models \cite{Babu:1988ig,Choudhury:1994vr,McDonald:2003zj}. We would
not attempt to review the topic but reemphasizing the point that for
a mechanism of radiative mass generation to be testable at colliders
\cite{Babu:2002uu,Chen:2006vn,AristizabalSierra:2006gb} the relevant
heavy mass scale cannot be too high.

The paper is organized as follows. We describe in some detail the
minimal model in the next section to set up our notations. The exact
constraints on the lepton masses and diagonalization matrices are
highlighted. They will be extensively utilized in our analytic
evaluation of radiative mass. Also listed are the Yukawa couplings
of leptons that may be useful in other applications. The radiative
mass is then calculated in section \ref{sec:analytic} in a manner
that facilitates later numerical analysis, and the final answer is
given in terms of some loop integrals. These integrals are defined
in Appendix A, and their leading terms in the heavy mass limit are
given. For numerical analysis in section \ref{sec:numerical}, we
first demonstrate the order of magnitude of radiative mass for a
heavy mass scale that would be accessible at colliders. Then we
consider the heavy mass limit trying to accommodate neutrino masses
derived from oscillation experiments. We conclude in the last
section where the main points of the work are recapitulated.

\section{Type III seesaw model}
\label{sec:model}

We describe systematically in this section the type III seesaw model
proposed in Ref. \cite{Foot:1988aq}. While the exposed relations
among the lepton mixing matrix and the lepton masses will be
employed in the next section to evaluate the radiative neutrino
masses, the displayed interactions may also be useful in other
applications.

\subsection{Yukawa couplings and lepton mass matrices}

The model introduces a lepton multiplet, $\Sigma$, that is a triplet
of $SU(2)_L$ but carries no hypercharge, on top of the fields
present in SM. We shall restrict ourselves to the leptonic sector of
the model. The lepton fields are
\begin{eqnarray}
F_L=\left(\begin{array}{c}n_L\\f_L\end{array}\right);~f_R;
~\Sigma_R=\left(\begin{array}{cc}\frac{1}{\sqrt{2}}\Sigma^0_R&\Sigma^+_R\\
\Sigma^-_R&-\frac{1}{\sqrt{2}}\Sigma^0_R\end{array}\right)
\end{eqnarray}
We have assumed without loss of generality that $\Sigma$ is
right-handed (RH). The Yukawa couplings plus the bare mass for
$\Sigma$ are
\begin{eqnarray}
-\calL_\text{Yuk}&=&\frac{1}{2}\text{tr}\left(
M_\Sigma\overline{\Sigma_R}\Sigma_R^C+%
M_\Sigma^*\overline{\Sigma_R^C}\Sigma_R\right)\nonumber\\
&&+\left(\overline{F_L}y_\Phi f_R\Phi+
\Phi^\dagger\overline{f_R}y^\dagger_\Phi F_L\right)
+\left(\overline{F_L}y_\Sigma\Sigma_R\tilde\Phi
+\tilde\Phi^\dagger\overline{\Sigma_R}y_\Sigma^\dagger F_L\right),
\end{eqnarray}
where $\Phi$ is the scalar doublet with
$\tilde\Phi=i\sigma^2\Phi^*$. $y_\Phi$ and $y_\Sigma$ are
respectively $3\times 3$ and $3\times 1$ complex Yukawa coupling
matrices. The superscript $C$ denotes the charge conjugation,
$\psi^C=\calC\gamma^0\psi^*$ with $\calC=i\gamma^0\gamma^2$. Our
notation is such that $\psi_L^C=(\psi_L)^C$. It is not necessary to
include a $F_L^C-\Sigma_R^C$ coupling since
$\overline{\psi^C}\chi^C=\bar\chi\psi$. Note that we can choose
$M_\Sigma$, which is the seesaw scale in the model, to be real
positive as any phase of it may be absorbed into $y_\Sigma$.

After $\Phi$ develops a vacuum expectation value, $v$, the lepton
mass terms become
\begin{eqnarray}
-\calL_\text{m}&=&\frac{1}{2}M_\Sigma\left(
\overline{\Sigma^0_R}\Sigma_R^{0C}+
\overline{\Sigma^-_R}\Sigma_R^{+C}+
\overline{\Sigma^+_R}\Sigma_R^{-C}+\text{h.c.}\right)\nonumber\\
&& +\frac{v}{\sqrt{2}}\left(\overline{f_L}y_\Phi f_R
+\frac{1}{\sqrt{2}}\overline{n_L}y_\Sigma\Sigma^0_R
+\overline{f_L}y_\Sigma\Sigma^-_R+\text{h.c.}\right)
\end{eqnarray}
Since $\Sigma_R^\pm$ carry electric charge, they cannot be Majorana
particles. Instead, their equal bare mass suggests the combination
to a Dirac field,
\begin{eqnarray}
\Psi=\Sigma_R^-+\Sigma_R^{+C}
\end{eqnarray}
with $\Sigma_R^{+C}=\calC\gamma^0(\Sigma_R^+)^*$. It is then
impossible to assign a lepton number to $\Psi$ without explicitly
breaking gauge symmetry. The lepton mass terms are summarized as
\begin{eqnarray}
-\calL_\text{m}&=&\frac{1}{2}\overline{N_L}m_NN_L^C+\overline{E_L}m_EE_R
+\text{h.c.}
\end{eqnarray}
where the neutral and charged lepton fields and their mass matrices
are
\begin{eqnarray}
&&N_L=\left(\begin{array}{c}n_L\\\Sigma_R^{0C}\end{array}\right),~~~
N_R=N_L^C=\left(\begin{array}{c}n_L^C\\\Sigma_R^0\end{array}\right),~~~
E=\left(\begin{array}{c}f\\\Psi\end{array}\right)
\label{eq:fields}\\
&&m_N=\left(\begin{array}{cc}0_3&\frac{1}{2}vy_\Sigma\\
\frac{1}{2}vy_\Sigma^T&M_\Sigma\end{array}\right),~~~
m_E=\left(\begin{array}{cc}
\frac{1}{\sqrt{2}}vy_\Phi&\frac{1}{\sqrt{2}}vy_\Sigma\\
0&M_\Sigma\end{array}\right)
\end{eqnarray}

\subsection{Gauge couplings of leptons}

The kinetic term for the triplet field is
\begin{eqnarray}
\calL^\Sigma_\text{kin}=\text{tr}\overline{\Sigma_R}i\Dslash\Sigma_R,
\end{eqnarray}
where the covariant derivative is
\begin{eqnarray}
D_\mu\Sigma_R=\partial_\mu\Sigma_R
-ig_2\frac{1}{2}[A^a_\mu\sigma^a,\Sigma_R]
\end{eqnarray}
with $A^a_\mu$ and $g_2$ being the $SU(2)_L$ gauge fields and
coupling. The kinetic term can be expressed in terms of the fields
defined in eq (\ref{eq:fields}). In so doing, the following
relations are useful,
$\overline{\psi^C}\gamma^\mu\chi^C=-\overline{\chi}\gamma^\mu\psi,~
\overline{\psi^C}\dslash\chi^C=\overline{\chi}\dslash\psi
-\partial_\mu(\overline{\chi}\gamma^\mu\psi)$, where the total
derivative may be dropped from Lagrangian.

Including the standard kinetic terms for the SM fields $F_L$ and
$f_R$, the complete kinetic terms for leptons are
\begin{eqnarray}
\calL_\textrm{kin}=\frac{1}{2}\bar Ni\dslash N+\bar Ei\dslash E%
+\frac{g_2}{\sqrt{2}}\left(J^{+\mu}_WW_{\mu}^++J^{-\mu}_WW_{\mu}^-\right)
+\frac{g_2}{c_W}J^\mu_ZZ_\mu+eJ^\mu_{\textrm{em}}A_\mu,
\end{eqnarray}
where $W^\pm_\mu,~Z_\mu$ and $A_\mu$ are the weak and
electromagnetic fields coupled to the currents
\begin{eqnarray}
J^{+\mu}_W&=&\Bar N\gamma^\mu(w_LP_L+w_RP_R)E\nonumber\\
J^\mu_Z&=&\bar N\gamma^\mu z_L^NP_LN
+\bar E\gamma^\mu(z_L^EP_L+z_R^EP_R)E\nonumber\\
J^\mu_\text{em}&=&-\bar E\gamma^\mu E %
\label{eq:currents1}
\end{eqnarray}
and $J^{-\mu}_W=(J^{+\mu}_W)^\dagger$, with the coupling matrices
being
\begin{eqnarray}
&&w_L=\left(\begin{array}{cc}1_3&\\
&\sqrt{2}\end{array}\right),~~~
w_R=\left(\begin{array}{cc}0_3&\\&\sqrt{2}\end{array}\right)
\nonumber\\
&&z_L^N=\left(\begin{array}{cc}\frac{1}{2}1_3&\\
&0\end{array}\right)\nonumber\\
&&z_L^E=\left(\begin{array}{cc}(-\frac{1}{2}+s_W^2)1_3&\\
&-c_W^2\end{array}\right),~~~
z_R^E=\left(\begin{array}{cc}s_W^21_3&\\
&-c_W^2\end{array}\right)
\end{eqnarray}
We have used the conventional notations $c_W=\cos\theta_W$,
$s_W=\sin\theta_W$, with $\theta_W$ being the weak angle.

\subsection{Diagonalization of lepton mass matrices}

Noting that the upper-left $3\times 3$ block of $m_N$ is zero, we
can make $m_N$ standardized as follows. A unitary transformation in
family space, $F_L\to\calU^\dagger F_L$, only modifies the Yukawa
couplings, $y_\Sigma\to\calU y_\Sigma$ and $y_\Phi\to\calU y_\Phi$.
One can choose $\calU$ to rotate the column vector $y_\Sigma$ to its
third component, so that
\begin{eqnarray}
m_N=\left(\begin{array}{ccc}%
0_2&&\\&0&\frac{1}{2}vr_\Sigma\\
&\frac{1}{2}vr_\Sigma&M_\Sigma
\end{array}\right)
\end{eqnarray}
where $r_\Sigma$ is real positive. There are thus two massless
neutrinos (named 1 and 2) at tree level. They will generally get a
radiative mass as their masslessness is not protected by any
symmetry. The other two neutrinos (3 and 4) get the masses
\begin{eqnarray}
m_{3,4}=\frac{1}{2}\left[\sqrt{M_\Sigma^2+(vr_\Sigma)^2}\mp
M_\Sigma\right]
\end{eqnarray}
The mass eigenstate fields of neutrinos are therefore
\begin{eqnarray}
\nu_L=U_N^TN_L,~\nu_R=\nu_L^C=U_N^\dagger N_R
\end{eqnarray}
where
\begin{eqnarray}
U_N=\left(\begin{array}{cc}1_2&\\&%
\begin{array}{rr}
ic_\theta&s_\theta\\-is_\theta&c_\theta\end{array}
\end{array}\right)
\end{eqnarray}
with $c_\theta=\cos\theta$, $s_\theta=\sin\theta$ and
$\tan\theta=\sqrt{m_3/m_4}$.

The mass matrix of the charged leptons is diagonalized by bi-unitary
transformations,
\begin{eqnarray}
E_{L,R}=U_{L,R}\ell_{L,R},~U_L^\dagger
m_EU_R=\text{diag}(m_e,m_\mu,m_\tau,m_\chi)
\end{eqnarray}
Here $\nu_4$ and $\chi$ are the new neutral and charged leptons
beyond SM. They must be very heavy to evade the experimental
detection so far. The tiny (small) mass of the observed neutrinos
(charged leptons) then implies that, to very good precision, we have
approximately
\begin{eqnarray}
m_4\approx m_\chi,~\theta^2\approx\frac{m_3}{m_4},
\label{eq:theta}
\end{eqnarray}
which will be employed in later numerical analysis.

\subsection{Summary of lepton interactions}

We can now express the interactions of leptons in terms of their
mass eigenstate fields, $\nu_i$ ($i=1,2,3,4$) and $\ell_\alpha$
($\alpha=e,\mu,\tau,\chi$). The currents in eq. (\ref{eq:currents1})
become
\begin{eqnarray}
J^{+\mu}_W&=&\Bar\nu\gamma^\mu(\calW_LP_L+\calW_RP_R)\ell
\nonumber\\
J^\mu_Z&=&\bar\nu\gamma^\mu\calZ^\nu_LP_L\nu+ \bar\ell\gamma^\mu
(\calZ^\ell_LP_L+\calZ^\ell_RP_R)\ell
\nonumber\\
J^\mu_\text{em}&=&-\bar\ell\gamma^\mu\ell%
\label{eq:currents2}
\end{eqnarray}
where
\begin{eqnarray}
&&\calW_L=U_N^Tw_LU_L,~\calW_R=U_N^\dagger w_RU_R
\nonumber\\
&&\calZ^\nu_L=U_N^Tz_L^NU_N^*
\nonumber\\
&&\calZ^\ell_L=U_L^\dagger z_L^EU_L,~
\calZ^\ell_R=U^\dagger_Rz_R^EU_R
\end{eqnarray}
Note that there is a degree of freedom in presenting the neutral
current of Majorana neutrinos. Using $\nu=\nu_L+\nu_L^C=\nu^C$ and
$\overline{\psi^C}\gamma^\mu P_{L,R}\chi^C=-\bar\chi\gamma^\mu
P_{R,L}\psi$, we can write
\begin{eqnarray}
\bar\nu\gamma^\mu\calZ^\nu_LP_L\nu
=\frac{1}{2}\bar\nu\gamma^\mu\left(\calZ^\nu_LP_L -\calZ^{\nu
T}_LP_R\right)\nu
\end{eqnarray}

Since $U_N$ and $w,~z$ are known, the following explicit results are
useful:
\begin{eqnarray}
&&\calW_L=\left(\begin{array}{cc}
1_2&\\
&\begin{array}{rr}
ic_\theta&-i\sqrt{2}s_\theta\\s_\theta&\sqrt{2}c_\theta
\end{array}
\end{array}\right)U_L,~~~
\calW_R=\sqrt{2}\left(\begin{array}{cc}
0_2&\\
&\begin{array}{rr} 0&is_\theta\\0&c_\theta
\end{array}
\end{array}\right)U_R
\nonumber\\
&&\calZ_L^\nu=\frac{1}{2}\left(\begin{array}{ccc} 1_2&&\\
&c^2_\theta&ic_\theta s_\theta\\
&-ic_\theta s_\theta&s_\theta^2
\end{array}\right)%
\label{eq:WZ}
\end{eqnarray}
One observes from the above that the right-handed charged current
involves only the massive neutrinos $\nu_{3,4}$ while the flavor
changing neutral currents occur for both charged leptons and
(massive) neutrinos.

For completeness, we present some additional results that may be
useful in other applications of the model. First of all, one can
construct the coupling matrices in the neutral currents in terms of
those in the charged currents:
\begin{eqnarray}
\calZ^\nu_L=1_4-\frac{1}{2}\calW_L\calW_L^\dagger,~
\calZ^\ell_L=s_W^21_4-\frac{1}{2}\calW_L^\dagger\calW_L,~
\calZ^\ell_R=s_W^21_4-\frac{1}{2}\calW_R^\dagger\calW_R
\end{eqnarray}
The Yukawa couplings of the would-be Goldstone bosons $G^{\pm,0}$
are
\begin{eqnarray}
\calL_\text{Yuk}^{G^{0,\pm}}
&=&+\frac{g_2}{\sqrt{2}m_W}G^+\bar\nu\left[m_\nu(\calW_LP_L+\calW_RP_R)
-(\calW_LP_R+\calW_RP_L)m_\ell\right]\ell+\text{h.c.}
\nonumber\\
&&-\frac{ig_2}{c_Wm_Z}G^0\bar\nu\left[m_\nu \calZ^\nu_L P_L
-\calZ^\nu_L m_\nu P_R\right]\nu
\nonumber\\
&&-\frac{ig_2}{c_Wm_Z}G^0\bar\ell\left[m_\ell(\calZ^\ell_LP_L+\calZ^\ell_RP_R)
-(\calZ^\ell_LP_R+\calZ^\ell_RP_L)m_\ell\right]\ell
\end{eqnarray}
where $m_\nu$ and $m_\ell$ are the diagonal mass matrices of the
neutrinos and charged leptons. The above simple structure is
dictated by the nature of $G^{\pm,0}$ although the intermediate
results in a direct derivation from $\calL_\textrm{Yuk}$ may look
cumbersome. In constrast, the Yukawa couplings to the physical Higgs
field $h$ are quite different since the leptons obtain masses from
both the bare mass term and the Yukawa couplings:
\begin{eqnarray}
\calL^h_\text{Yuk}&=&-\frac{h}{v}m_3c_\theta^2
\left(\overline{\nu_3}\nu_3+\overline{\nu_4}\nu_4\right) %
-i\frac{h}{v}(m_4-m_3)c_\theta s_\theta
\left(\overline{\nu_{3L}}\nu_{4R}-\overline{\nu_{4R}}\nu_{3L}\right)
\nonumber\\
&&-\frac{h}{v}\left\{\overline{\ell_{\alpha L}}
\left[m_\alpha\delta_{\alpha\beta}-(m_4-m_3)U^*_{L\chi\alpha}
U_{R\chi\beta}\right]\ell_{\beta R}+\textrm{h.c.}\right\}
\end{eqnarray}

\subsection{Constraints on mixing matrices and lepton masses}

For convenience in the next section, we collect here the constraints
on $U_{L,R}$, $m_\alpha$ and $m_i$:
\begin{eqnarray}
C1&:&m_3^2c_\theta=m_4^2s_\theta\\
C2&:&\sum_\alpha U^*_{Li\alpha}U_{L3\alpha}=\sum_\alpha
U^*_{Li\alpha}U_{L4\alpha}=0\\
C3&:&\sum_\alpha m_\alpha
U_{Li\alpha}^*U_{R4\alpha}=0\\
C4&:&\sum_\alpha m_\alpha^2U_{Li\alpha}^*U_{L4\alpha}=0
\end{eqnarray}
where $i=1,~2$ in $C2,~C3$ and $C4$. They will be extensively used
to improve the apparent convergence of the loop integrals and
extract the leading terms in the large mass limit of heavy leptons.
These constraints are exact and can be readily derived. The
constraint $C1$ is from diagonalization of $m_N$ while $C2$
represents unitarity of $U_L$. After rotating the column vector
$y_\Sigma$ to its third component, the first two columns in the last
row of $m_E^\dagger$ vanish. This yields $(U_Rm_\ell^\dagger
U_L^\dagger)_{4i}=(m_E^\dagger)_{4i}=0$ for $i=1,2$, which is $C3$.
In addition, we find that $(m_Em_E^\dagger)_{4i}$ also vanishes for
$i=1,~2$, which gives the last constraint $C4$. For the sake of
notational simplicity, we sometimes also use the Latin letters
$i,~j$ and numbers, which enter through the charged current matrices
$\calW_{L,R}$, as the indices for the corresponding charged leptons.

\section{Two-loop induced neutrino masses}
\label{sec:analytic}

Now we calculate the radiative mass of the neutrinos $\nu_{1,2}$
that are massless at tree level. This is given by their minus
self-energy evaluated at the zero momentum. We thus need to
calculate the amplitude for the transition, $\nu_{iL}\to\nu_{jL}^C$,
with $i,~j=1,~2$. There is no contribution at one loop. This arises
because, while the neutral current does not couple $\nu_{1,2}$ to
the massive ones $\nu_{3,4}$, the charged current involving
$\nu_{1,2}$ is purely left-handed and thus cannot induce a mass for
a massless particle.

At two loops, we note first that a diagram with at least one of the
two external lines connected to a virtual $Z$ boson cannot
contribute. This is because, if it did, removing this virtual $Z$
line would also do since $Z$ couples diagonally to $\nu_{1,2}$ and
conserves chirality. But this would contradict our claim at one
loop. The external lines must therefore all connect to virtual
$W^\pm$ bosons. Finally, the two external lines cannot connect to
the same virtual $W^\pm$ due to charge conservation. This leaves
with us the single diagram shown in Fig. 1.
\begin{center}
\begin{picture}(200,80)(0,0)
\SetOffset(20,45) %
\ArrowLine(-20,0)(0,0)\ArrowLine(0,0)(40,0)\Line(40,0)(60,0)
\ArrowLine(120,0)(100,0)\ArrowLine(100,0)(60,0)%
\PhotonArc(30,0)(30,0,180){3}{11}\PhotonArc(70,0)(30,180,360){3}{11}
\Text(-10,-8)[]{$\nu_{iL}$}\Text(20,8)[]{$\ell_\alpha$}
\Text(50,-8)[]{$\nu_k$}\Text(80,8)[]{$\ell_\beta$}
\Text(110,-8)[]{$\nu_{jL}$}
\Text(5,25)[r]{$W^-$}\Text(95,-25)[l]{$W^-$}%
\Text(20,-8)[]{$p$}\Text(80,-8)[]{$q$}%
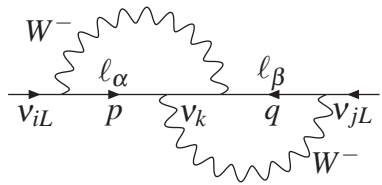
\Text(60,-50)[]{Figure 1. Diagram contributing to $-i\Sigma_{ji}$}

\end{picture}
\end{center}

We shall evaluate the radiative mass in unitarity gauge. We first
simplify and classify the contributions from the diagram. Then we
apply the constraints $C1-C4$ to reach manifest convergence in loop
integrals and to get prepared for isolating leading terms in the
seesaw limit. Finally, the contributions are expressed in terms of
some standard parameter integrals.

To start with, we note that the external $\nu_{i,j}$ ($i,~j=1,~2$)
have no right-handed couplings to the corresponding virtual charged
leptons $\ell_{\alpha,\beta}$. The diagram then decomposes into four
terms according to the chiralities of the two vertices involving the
virtual neutrino $\nu_k$. After some algebraic work, we can remove
all $\gamma$ matrices in favor of the products of loop momenta and
obtain
\begin{eqnarray}
u^T_j\Sigma_{ji}u_i&=&\calM_{ji}u^T_j\calC P_Lu_i
\nonumber\\
\calM_{ji}&=&\frac{g_2^4}{4(4\pi)^4}\left[
T^{LL}+T^{RR}+T^{RL}+T^{RL}|_{i\leftrightarrow j}\right]
\end{eqnarray}
where $u_{i,j}$ are the spinors for external neutrinos, and $\calM$
gives the radiative neutrino mass. The $T$ functions are
\begin{eqnarray}
T^{LL}&=&m_k\calW^*_{Li\alpha}\calW^*_{Lj\beta}
\calW_{Lk\alpha}\calW_{Lk\beta}F^{LL}(\alpha,\beta;k)
\nonumber\\
T^{RR}&=&m_km_\alpha m_\beta
m_W^{-2}\calW^*_{Li\alpha}\calW^*_{Lj\beta}
\calW_{Rk\alpha}\calW_{Rk\beta}F^{RR}(\alpha,\beta;k)
\nonumber\\
T^{RL}&=&m_\alpha\calW^*_{Li\alpha}\calW^*_{Lj\beta}
\calW_{Rk\alpha}\calW_{Lk\beta}F^{RL}(\alpha,\beta;k)
\end{eqnarray}
where the loop functions $F$ are dimensionless functions of the mass
ratios. Upon Wick rotation to Euclidian space, they become
\begin{eqnarray}
F^{LL}(\alpha,\beta;k)&=&-\iint\frac{p\cdot q}{D(\alpha,\beta;k)}
\left[4+p^2q^2+4(p^2+q^2)\right]
\nonumber\\
F^{RR}(\alpha,\beta;k)&=&+\iint\frac{1}{D(\alpha,\beta;k)}\left[-8%
+2(p\cdot q)^2-q^2p^2-2(p^2+q^2)\right]
\nonumber\\
F^{RL}(\alpha,\beta;k)&=&+\iint\frac{1}{D(\alpha,\beta;k)} \left[%
-4(p\cdot q+q^2)-p^2q^2(p\cdot q+q^2)\right.
\nonumber\\
&&\left.+2(p^2p\cdot q+2(p\cdot q)^2-q^2p^2) -4q^2(p\cdot
q+q^2)\right]
\end{eqnarray}
where the notations are
\begin{eqnarray}
&&D(\alpha,\beta;k)=[(p+q)^2+r_k][p^2+r_\alpha][p^2+1]
[q^2+r_\beta][q^2+1]
\nonumber\\
&&r_k=\frac{m^2_k}{m^2_W},~r_\alpha=\frac{m_\alpha^2}{m_W^2},
~\iint=\frac{1}{\pi^4}\int d^4p\int d^4q
\end{eqnarray}
Here the summation over the virtual lepton flavors
$\alpha,~\beta,~k$ is implied in the $T$ functions, and $\calM_{ji}$
is manifestly symmetric as expected for Majorana particles.

Since only massive neutrinos enter the right-handed charged current,
the virtual $\nu_k$ in the $T$ functions is actually restricted to
$\nu_{3,4}$. Using the explicit forms of $\calW_{L,R}$ shown in eq.
(\ref{eq:WZ}), the $T$ functions decompose into
\begin{eqnarray}
T^{LL}&=&U^*_{Li\alpha}U^*_{Lj\beta}\Big\{
U_{L3\alpha}U_{L3\beta}\left[m_4s_\theta^2F^{LL}_4
-m_3c_\theta^2F^{LL}_3\right]
\nonumber\\
&& +2U_{L4\alpha}U_{L4\beta}\left[m_4c_\theta^2F^{LL}_4
-m_3s_\theta^2F^{LL}_3\right]
\nonumber\\
&&+\sqrt{2}c_\theta s_\theta
(U_{L4\alpha}U_{L3\beta}+U_{L3\alpha}U_{L4\beta})
\left[m_4F^{LL}_4+m_3F^{LL}_3\right]\Big\}
\nonumber\\
T^{RR}&=&2m_\alpha m_\beta m_W^{-2}U^*_{Li\alpha}U^*_{Lj\beta}
U_{R4\alpha}U_{R4\beta}\left[m_4c_\theta^2F^{RR}_4
-m_3s_\theta^2F^{RR}_3\right]
\nonumber\\
T^{RL}&=&m_\alpha U^*_{Li\alpha}U^*_{Lj\beta}U_{R4\alpha}%
\Big\{\sqrt{2}c_\theta s_\theta U_{L3\beta}
\left[F^{RL}_4-F^{RL}_3\right] %
+2U_{L4\beta}\left[s_\theta^2F^{RL}_3+
c_\theta^2F^{RL}_4\right]\Big\} %
\label{eq:Tfunc}
\end{eqnarray}
where for brevity the first two arguments $\alpha,~\beta$ of the $F$
functions are suppressed while the third one $k$ appears as a
subscript $3$ or $4$. In addition to improving apparent convergence,
the main merit of applying the constraints $C1-C4$ is to subtract
heavy leptons $\nu_4,~\chi$ from the loops. This avoids manifestly
in the contributing terms some large numbers that are actually
balanced by the small matrix elements mixing the light and heavy
leptons. Furthermore, this facilitates the extraction of the leading
terms that can survive upon being multiplied by the mixing matrix
elements and summing over light flavors $\alpha,~\beta$, for which
the hierarchical limit $1\gg r_\alpha\gg r_3$ works very well. We
stress that we are not discarding the contributions from heavy
leptons but are combining them in a judicious manner with those from
light leptons before numerical analysis is done. In the following
subsections we shall reduce the $T$ functions using the constraints.

\subsection{Reduction of $T^{LL}$}

We note first of all that the numerator of $F^{LL}$ is separately
linear in $p^2$ and $q^2$. Take $p^2$ as an example. By decomposing
$p^2=(p^2+r_\alpha)-r_\alpha$, the first term cancels the
corresponding factor in $D(\alpha,\beta;k)$ so that its contribution
to $F^{LL}$ is independent of $\alpha$. The constraint $C2$ then
implies that it does not survive in $T^{LL}$ upon summing over
$\alpha$. We can thus effectively set in the numerator of $F^{LL}$,
$p^2\to -r_\alpha$ and similarly $q^2\to -r_\beta$:
\begin{eqnarray}
F^{LL}(\alpha,\beta;k)&\to&-[4+r_\alpha r_\beta-4(r_\alpha+r_\beta)]
\iint\frac{p\cdot q}{D(\alpha,\beta;k)},
\end{eqnarray}
where the arrow means equality when multiplied by $U$ factors and
summing over $\alpha,~\beta$. To go further, we have to cope
separately with the four terms in $T^{LL}$ according to the $U_L$
factors involved:
\begin{eqnarray}
T^{LL}&=&U^*_{Li\alpha}U^*_{Lj\beta}\Big\{
U_{L3\alpha}U_{L3\beta}T^{LL}_{33}+2U_{L4\alpha}U_{L4\beta}T^{LL}_{44}
\nonumber\\
&&+\sqrt{2}\left[U_{L4\alpha}U_{L3\beta}T^{LL}_{43}+U_{L3\alpha}
U_{L4\beta}T^{LL}_{34}\right]\Big\}
\end{eqnarray}
with obvious definitions on $T^{LL}_{33}$ etc by comparing with eq.
(\ref{eq:Tfunc}).

Although the first term, $T^{LL}_{33}$, is already convergent upon
applying $C1$ due to the subtraction between $F^{LL}_4$ and
$F^{LL}_3$, we can do better by subtracting explicitly the
contribution from the heavy charged lepton $\chi$. The trick is
that, for a term in $F^{LL}$ that is not proportional to $r_\alpha$
we make the substitution
\begin{eqnarray}
\frac{1}{p^2+r_\alpha}\to\frac{1}{p^2+r_\alpha}-\frac{1}{p^2+r_4}
\equiv d_\alpha(p)%
\label{eq:sub1}
\end{eqnarray}
while for a term that is proportional to $r_\alpha$, we do as
follows
\begin{eqnarray}
\frac{r_\alpha}{p^2+r_\alpha}\to\frac{r_\alpha}{p^2+r_\alpha}
-\frac{r_4}{p^2+r_4}\equiv e_\alpha(p)%
\label{eq:sub2}
\end{eqnarray}
The legitimacy of the substitutions is guaranteed by the constraint
$C2$. Thus,
\begin{eqnarray}
T^{LL}_{33}&\to&m_3c_\theta^2\iint\frac{p\cdot q}
{[p^2+1][q^2+1]}d_3(p+q)\nonumber\\
&&\times\left[4d_\alpha(p)d_\beta(q) +e_\alpha(p)e_\beta(q)
-4e_\alpha(p)d_\beta(q)-4d_\alpha(p)e_\beta(q)\right]
\end{eqnarray}

The second term, $T^{LL}_{44}$, is multiplied by
$U_{L4\alpha}U_{L4\beta}$ so that we have a choice of whether to use
the constraint $C2$ (i.e., eq. (\ref{eq:sub1})) or $C4$ (eq.
(\ref{eq:sub2})) for the terms proportional to $r_\alpha$ or
$r_\beta$. It turns out that the latter is better as it can reduce
the amount of work by bringing down more factors of
$r_{\alpha,\beta}$ for light leptons $\alpha,~\beta$, which makes
the corresponding term subdominant in the hierarchical limit. The
last two terms may be similarly manipulated. The results are
summarized as follows:
\begin{eqnarray}
T^{LL}_{44}&\to&[4+r_\alpha r_\beta-4(r_\alpha+r_\beta)]
\iint\frac{p\cdot q}{[p^2+1][q^2+1]}d_\alpha(p)d_\beta(q)
\nonumber\\
&&\times\left[\frac{m_3s_\theta^2}{(p+q)^2+r_3}
-\frac{m_4c_\theta^2}{(p+q)^2+r_4}\right] %
\nonumber\\
T^{LL}_{43}&\to&\iint\frac{p\cdot q}{[p^2+1][q^2+1]}
\left[4(r_\alpha-1)d_\beta(q)+(4-r_\alpha)e_\beta(q)\right]
d_\alpha(p)\nonumber\\
&&\times\left[\frac{m_4}{(p+q)^2+r_4}+\frac{m_3}{(p+q)^2+r_3}\right]
c_\theta s_\theta
\end{eqnarray}
while $T^{LL}_{34}$ is obtained from $T^{LL}_{43}$ by
$\alpha\leftrightarrow\beta$ and $i\leftrightarrow j$. Since
$\alpha,~\beta$ are summed over, this amounts to symmetrizing
$T^{LL}_{43}$ in $i,~j$.

The advantage of the above results can be understood by recalling
that we now only need to sum over light flavors $\alpha,~\beta$ in
$T^{LL}$. Since $1\gg r_{\alpha,\beta}\gg r_3$, it is numerically
very good to set $r_\alpha=r_\beta=r_3=0$. For instance, the largest
$r_\tau\sim 5\times 10^{-4}$ while $r_3\sim 6\times 10^{-24}$ for
$m_3\sim 0.2~\eV$. This will not introduce mass singularities in the
loop integrals. In addition, when a term proportional to $m_3$ is
accompanied by one proportional to $m_4$, we ignore the former since
it cannot make a significant contribution to the radiative mass.
(Note that $T^{LL}$ is exceptional since
$m_3c_\theta^2=m_4s_\theta^2$.) Although the above argument is
self-evident, we have inspected and compared carefully all of the
terms to verify it. This simplifies considerably the integrals to
compute:
\begin{eqnarray}
T^{LL}_{33}&\to&m_3c_\theta^2r_4^2
\left\{4\left[\calX_2(0)-\calX_2(r_4)\right]
+8\left[\calX_1(0)-\calX_1(r_4)\right]+\calX_0\right\}
\nonumber\\
T^{LL}_{44}&\to&-m_4c_\theta^24r_4^2\calX_2(r_4)
\nonumber\\
T^{LL}_{43}&\to&-m_4c_\theta s_\theta
4r_4^2[\calX_2(r_4)+\calX_1(r_4)]%
\label{eq:TLL}
\end{eqnarray}
and $T^{LL}_{34}=T^{LL}_{43}$, where the loop integrals $\calX$ are
defined in Appendix A. These functions are independent of
$\alpha,~\beta$ and depend only on $r_4$.

\subsection{Reduction of $T^{RR}$}

The second term in $T^{RR}$ is doubly suppressed by $m_3s_\theta^2$
compared to the first one and will be ignored from the start. Since
the numerator in the integrand of $F^{RR}_4$ is again linear in
$p^2$ and $q^2$, they may be replaced by $-r_\alpha$ and $-r_\beta$
respectively employing the constraint $C3$. For the $2(p\cdot q)^2$
term in the numerator, we decompose as follows,
\begin{eqnarray*}
2(p\cdot q)^2=p\cdot q\left([(p+q)^2+r_4]-[p^2+r_\alpha]
-[q^2+r_\beta]+[r_\alpha+r_\beta-r_4]\right)
\end{eqnarray*}
The first term is cancelled by the same factor in the denominator
$D$ making the integrand odd in $p$, and thus vanishes upon
integration. The second term again cancels a same factor from $D$
and is killed upon summing over $\alpha$ by the constraint $C3$, and
the same happens with the third term as well. The numerator now
becomes effectively,
\begin{eqnarray*}
-(8+r_4p\cdot q)+(p\cdot q+2)(r_\alpha+r_\beta)-r_\alpha r_\beta
\end{eqnarray*}

Now we make the substitutions in eqs.(\ref{eq:sub1},\ref{eq:sub2})
as we did in the previous subsection, though employing this time the
constraint $C3$, to obtain,
\begin{eqnarray}
F^{RR}_4&\to&\iint\frac{1}{[p^2+1][q^2+1][(p+q)^2+r_4]}
\left\{-(8+r_4p\cdot q)d_\alpha(p)d_\beta(q)\right.
\nonumber\\
&&\left.+(p\cdot q+2)\left[e_\alpha(p)d_\beta(q)
+e_\beta(p)d_\alpha(q)\right]-e_\alpha(p)e_\beta(q)\right\}
\end{eqnarray}
Since it is now legitimate to sum only over light flavors
$\alpha,~\beta$, the above simplifies to
\begin{eqnarray}
F^{RR}_4\to -r_4^2\Big\{8\calY_2(r_4)+4\calY_1(r_4)+\calY_0(r_4)
+r_4\calX_2(r_4)+2\calX_1(r_4)\Big\}%
\label{eq:FRR4}
\end{eqnarray}
where the new integrals $\calY$ are also defined in Appendix A.

\subsection{Reduction of $T^{RL}$}

This chirality-mixed part from the two vertices involving the
virtual neutrino $\nu_k$ contains the most number of terms in
$F^{RL}_k$:
\begin{eqnarray}
T^{RL}=m_\alpha U^*_{Li\alpha}U^*_{Lj\beta}U_{R4\alpha}%
\Big\{\sqrt{2}c_\theta s_\theta U_{L3\beta}
\left[F^{RL}_4-F^{RL}_3\right]+2c_\theta^2U_{L4\beta}F^{RL}_4\Big\}
\end{eqnarray}
where we have dropped the $s_\theta^2F^{RL}_3$ term as one cannot
rely on it to induce a reasonable mass due to a tiny $s_\theta^2\sim
10^{-12}$ at $m_3\sim 0.2~\eV$ and $m_4\sim 200~\GeV$, for instance.

The numerator of the integrand in $F^{RL}$ is linear in $p^2$, which
can thus be replaced by $-r_\alpha$ using the constrain $C3$. On the
other hand, since the numerator is quadratic in $q^2$, we must
distinguish between the two terms in $T^{RL}$ which are proportional
to $U_{L3\beta}$ and $U_{L4\beta}$ respectively. For the first one,
we can only set one factor of $q^2$ to $-r_\beta$ using $C2$. After
this, we apply $C2$ and $C3$ via the substitutions in eqs.
(\ref{eq:sub1},\ref{eq:sub2}) and obtain,
\begin{eqnarray}
F^{RL}_4-F^{RL}_3&\to& %
\iint\frac{d_3(p+q)}{[p^2+1][q^2+1]}
\left[(p\cdot q+q^2+2)e_\alpha(p)e_\beta(q)%
+2p\cdot qe_\alpha(p)d_\beta(q)\right.
\nonumber\\%
&&\left.-4(p\cdot q+q^2+1)d_\alpha(p)e_\beta(q)%
+4\left(p\cdot q-(p\cdot q)^2\right)d_\alpha(p)d_\beta(q)\right]
\end{eqnarray}
The summation over light flavors $\alpha,~\beta$ then yields the
result in terms of the standard integrals:
\begin{eqnarray}
F^{RL}_4-F^{RL}_3&\to&r_4^2\Big\{
r_4[\calU_0+4\calU_1]+\calX_0+6[\calX_1(0)-\calX_1(r_4)]
\nonumber\\
&&+4[\calX_2(0)-\calX_2(r_4)]
-2r_4\calX_2(r_4)+[\calY_0(0)-\calY_0(r_4)]\Big\}%
\label{eq:FRL43}
\end{eqnarray}

For the second term proportional to $U_{L4\beta}$ in $T^{RL}$, we
can set two factors of $q^2$ to $-r_\beta$ because of $C2$ and $C4$.
The subsequent manipulation based on the constraints and eqs.
(\ref{eq:sub1},\ref{eq:sub2}) is similar, and gives
\begin{eqnarray}
F^{RL}_4&\to&r_4^2\Big\{r_\beta\left(\calY_0(r_4)+[\calX_1(r_4)
+2\calY_1(r_4)]+4\calY_1(r_4)+4[\calY_2(r_4)+\calX_2(r_4)]\right)
\nonumber\\
&&-4\calX_2(r_4)-2[\calX_1(r_4)+r_4\calX_2(r_4)]\Big\}%
\label{eq:FRL4}
\end{eqnarray}
where the terms suppressed by $r_\beta$ will be ignored from now on.

To finish this section, we summarize the terms in the radiative
neutrino mass as follows:
\begin{eqnarray}
\calM_{ji}&=&\frac{m_W^4G_F^2}{2^5\pi^4}U^*_{Li\alpha}U^*_{Lj\beta}
\nonumber\\
&\times&\Big\{U_{L3\alpha}U_{L3\beta}T^{LL}_{33}
+2U_{L4\alpha}U_{L4\beta}T^{LL}_{44}
+\sqrt{2}\left(U_{L4\alpha}U_{L3\beta}+U_{L3\alpha}
U_{L4\beta}\right)T^{LL}_{43}
\nonumber\\
&&+2\sqrt{r_\alpha r_\beta}
U_{R4\alpha}U_{R4\beta}m_4c_\theta^2F^{RR}_4
\nonumber\\
&&+\sqrt{2}c_\theta s_\theta \left(%
m_\alpha U_{R4\alpha}U_{L3\beta}+m_\beta
U_{R4\beta}U_{L3\alpha}\right)\left(F^{RL}_4-F^{RL}_3\right)
\nonumber\\
&&+2c_\theta^2\left(m_\alpha U_{R4\alpha}U_{L4\beta} +m_\beta
U_{R4\beta}U_{L4\alpha}\right)F^{RL}_4\Big\}%
\label{eq:mass}
\end{eqnarray}
where the relevant functions are given in eqs. (\ref{eq:TLL},
\ref{eq:FRR4}, \ref{eq:FRL43}, \ref{eq:FRL4}) in terms of the
standard integrals calculated in Appendix A. The summation over the
light charged leptons $\ell_\alpha$ and $\ell_\beta$ is understood
in the above.

\section{Numerical analysis}
\label{sec:numerical}

Now we investigate whether we can accommodate the neutrino masses
measured in oscillation experiments. Our starting formula was given
in (\ref{eq:mass}) which involves the light-heavy mixing parameters
in addition to the upper-left $3\times 3$ submatrix of $U_L$. From
eq. (\ref{eq:WZ}) we see that the latter is just the leptonic mixing
matrix measured in oscillation experiments to very good precision.
However it is no more exactly unitary, and the deviation from
unitarity is determined by the light-heavy mixing. A realistic
numerical estimate should take all this into account to avoid a
misleading conclusion. Although a global fitting to the lepton
mixing parameters is possible with radiative corrections included,
our main result on the seesaw scale required to reproduce the
neutrino masses is independent of this fitting.

Both matrices $m_Em_E^\dagger$ and $m_E^\dagger m_E$ for the charged
leptons have the hierarchical structure
\begin{eqnarray}
M=\left(\begin{array}{cc}%
B&d\\d^\dagger&A
\end{array}\right),
\end{eqnarray}
where $B$ and $d$ are respectively a $3\times 3$ and $3\times 1$
matrix, whose entries are much smaller in magnitude than the
positive number $A$. Then, the submatrix of the diagonalization
matrix that mixes the small and large entries can be estimated as
$\kappa\approx dA^{-1}$. Application of this to $m_Em_E^\dagger$ and
$m_E^\dagger m_E$ yields for $\alpha=e,~\mu,~\tau$:
\begin{eqnarray}
U_{L4\alpha}&\sim&(m_3/m_4)^{1/2}=(r_3/r_4)^{1/4}=\theta
\nonumber\\
U_{R4\alpha}&\sim&(m_3/m_4)^{1/2}(m_\alpha/m_4)
=\theta(r_\alpha/r_4)^{1/2}%
\label{eq:U4}
\end{eqnarray}
where (\ref{eq:theta}) is used. And the unitarity violation in the
submatrix of light leptons is, for $i=1,~2$,
\begin{eqnarray}
\sum_{\alpha=e,\mu,\tau}U^*_{Li\alpha}U_{L3\alpha}
=-U^*_{Li\chi}U_{L3\chi}\sim\theta^2%
\label{eq:violation}
\end{eqnarray}

Consider first the case in which $m_4$ is not very large. This is
the range of parameters that is particularly relevant to LHC
physics. A heavy active lepton, especially the charged one $\chi$,
is supposed to be accessible if it is not much heavier than several
hundred GeV. Our estimate of heavy-light mixing parameters is still
good enough since $m_4$ is much larger than the light lepton masses.
Using the estimates in eqs.(\ref{eq:theta}, \ref{eq:U4}) (but not
yet the one in (\ref{eq:violation})), we find that the three classes
of contributions to $\calM_{ji}$ in eq. (\ref{eq:mass}) consist of
the following terms in units of $2^{-5}\pi^{-4}m_W^4G_F^2m_3$:
\begin{eqnarray}
LL&:&U^*_{Li\alpha}U^*_{Lj\beta}U_{L3\alpha}U_{L3\beta},
~U^*_{Li\alpha}U^*_{Lj\beta},
~\big(U^*_{Li\alpha}U^*_{Lj\beta}+U^*_{Lj\alpha}U^*_{Li\beta}\big)
U_{L3\beta}\nonumber\\
RR&:&r_\alpha r_\beta U^*_{Li\alpha}U^*_{Lj\beta}
\nonumber\\
RL&:&r_\alpha(U^*_{Li\alpha}U^*_{Lj\beta}+U^*_{Lj\alpha}U^*_{Li\beta})
U_{L3\beta},
~r_\alpha(U^*_{Li\alpha}U^*_{Lj\beta}+U^*_{Lj\alpha}U^*_{Li\beta})
\end{eqnarray}
where each term is to be multiplied by a coefficient that is a sum
of integrals as can be obtained from eqs. (\ref{eq:TLL},
\ref{eq:FRR4}, \ref{eq:FRL43}, \ref{eq:FRL4}). The point is that
these coefficients are order one numbers for $r_4$ not very large.
Then, independently of the mixing matrix of light leptons, it is
safe to say that
\begin{eqnarray}
|\calM_{ji}|<1.8\times 10^{-6}m_3
\end{eqnarray}
Since no light neutrinos can be heavier than an eV from cosmological
considerations, there is no hope to induce a large enough radiative
mass $m_1$ or $m_2$ from $m_3$. Therefore, the minimal type III
seesaw model cannot accommodate oscillation data if the heavy
leptons have an intermediate mass. To put another way, the
oscillation data already excludes the possibility that the active
heavy leptons in the model would be accessible at LHC.

It is interesting to ask whether there is a chance at all in the
model to induce a large enough neutrino mass. For this purpose, we
study the seesaw limit in which $m_4$ blows up. Then $\calM_{ji}$ is
a sum of the following terms (again in units of
$2^{-5}\pi^{-4}m_W^4G_F^2m_3$):
\begin{eqnarray}
LL&:&r_3r_4\left[\calX_0\right]
-U^*_{Li\alpha}U^*_{Lj\beta}8[r_4^2\calX_2(r_4)]
-\big(U^*_{Li\alpha}+U^*_{Lj\alpha}\big)
4\sqrt{2r_3r_4}[r_4\calX_1(r_4)]
\nonumber\\
RR&:&-r_\alpha U^*_{Li\alpha}r_\beta U^*_{Lj\beta}2\left[
r_4\calY_0(r_4)+r_4^2\calX_2(r_4)+2r_4\calX_1(r_4)\right]
\nonumber\\
RL&:&\left(r_\alpha U^*_{Li\alpha}\sqrt{2r_3r_4}
\left[r_4\calU_0+\calX_0\right]%
-r_\alpha U^*_{Li\alpha}U^*_{Lj\beta}4\left[r_4\calX_1(r_4)
+r_4^2\calX_2(r_4)\right]\right)
\nonumber\\
&&+(i\leftrightarrow j)%
\label{eq:large1}
\end{eqnarray}
All combinations of loop integrals in the square brackets are $O(1)$
constants up to logarithmic corrections in the large $r_4$ limit. We
have also taken into account the unitarity violation estimated in
eq. (\ref{eq:violation}). Because of the estimates employed, the
relative sign and factors of two between terms in the above cannot
be taken seriously. But this does not preclude us from making a
definite conclusion as shown below.

To induce a mass of $O(m_3)$, some terms in eq. (\ref{eq:large1})
must be above $10^5$. This obviously requires a large $r_4$. But
even this is insufficient. On the one hand, the terms not multiplied
by $r_4$ factors outside the square brackets can be safely ignored;
on the other, all remaining terms are controlled by $r_3r_4$. We
must therefore require $r_3r_4\gg 1$. This corresponds to the
combined limit in terms of the original parameters in Lagrangian,
$M_\Sigma\gg vr_\Sigma\gg m_W$. In the limit, only the first term in
the $LL$ class is relevant:
\begin{eqnarray}
\calM_{ji}\sim 2^{-5}\pi^{-4}m_W^4G_F^2m_3r_3
r_4\left[\calX_0\right]
\end{eqnarray}
Inspection of our derivation shows that this is the term that is
doubly suppressed by unitarity violation between the third row and
the first two rows of the light lepton mixing matrix. But
unfortunately it is impractical to measure the violation down to the
level that we are interested in, i.e.,
$\sim\theta^2=r_3/r_4=m_3^2/m_4^2$. The information on the indices
$(i,j)$ is lost also because of the estimates employed. This means
in passing that our analysis on the neutrino masses in the above
limit is independent of a detailed fitting to the leptonic mixing
parameters. We find it is natural for the model to favor the normal
hierarchy scenario; namely, a larger $m_3$ seeds a smaller
$m_{1,2}$. For the purpose of illustration, we assume $m_1=0$. The
solar and atmospherical oscillation data then give $m_2\approx
8.7\times 10^{-3}~\eV$ and $m_3\approx 4.9\times 10^{-2}~\eV$
respectively, which can be fulfilled by requiring
\begin{eqnarray}
m_4\sim 4\times 10^{16}~\GeV
\end{eqnarray}
This is roughly the scale of grand unification.

\section{Conclusion}

The minimal type III seesaw model introduces a lepton triplet on top
of the particles in SM. Two neutrinos out of four are massless at
the tree level, but they are not protected by any symmetry from
getting a radiative mass at the quantum level. We have shown that
the latter takes place first at two loops, and determined it in
terms of some parameter functions. By employing realistic estimates
of the mixing parameters between the light and heavy leptons, we
studied the pattern of the neutrino masses. We found that it is not
possible to accommodate the spectrum determined in oscillation
experiments if the heavy leptons have a mass that would be within
the reach of LHC. However, if the seesaw scale is as large as that
of grand unification, it is still possible to accommodate the
spectrum in a nice manner: one light neutrino gets mass directly
from seesaw while the other two get a radiative mass. The model
would then contain nothing new but the tiny neutrino masses. The
main message extracted from this work is therefore, if LHC sees
something like a triplet lepton, it definitely comes from a
structure that goes beyond the economical one as originally
suggested.

\vspace{0.5cm}
\noindent %
{\bf Acknowledgement} This work is supported in part by the grants
NCET-06-0211 and NSFC-10775074.

{\Large\bf Appendix A: Loop integrals}
\\

The loop integrals in the final result of $T^{LL}$ (see
eq.(\ref{eq:TLL})) are defined as
\begin{eqnarray}
\calX_2(r)&=&\iint\frac{p\cdot q}{D_1(r)p^2q^2}
\nonumber\\
\calX_1(r)&=&\iint\frac{p\cdot q}{D_1(r)q^2}
\nonumber\\
\calX_0&=&\iint\frac{p\cdot qr_4}{D_1(r_4)(p+q)^2}
\end{eqnarray}
where
\begin{eqnarray}
D_1(r)=[p^2+r_4][p^2+1][q^2+r_4][q^2+1][(p+q)^2+r]
\end{eqnarray}
The new integrals appearing in $T^{RR}$ and $T^{RL}$ are
respectively,
\begin{eqnarray}
\calY_2(r)&=&\iint\frac{1}{D_1(r)p^2q^2}
\nonumber\\
\calY_1(r)&=&\iint\frac{1}{D_1(r)q^2}
\nonumber\\
\calY_0(r)&=&\iint\frac{1}{D_1(r)}
\end{eqnarray}
and
\begin{eqnarray}
\calU_0&=&\iint\frac{1}{D_2}
\nonumber\\
\calU_1&=&\iint\frac{1} {D_2q^2}
\end{eqnarray}
where
\begin{eqnarray}
D_2=(p+q)^2[(p+q)^2+r_4][q^2+1][q^2+r_4][p^2+r_4]
\end{eqnarray}
There is another integral in calculating $T^{RL}$ that can be
related to those already defined:
\begin{eqnarray*}
\iint\frac{(p\cdot q)^2}{D_1(r)p^2q^2}
=-\calX_1(r)-\frac{r}{2}\calX_2(r)
\end{eqnarray*}

The basic technique to compute the above integrals is to use
fractions and the one-loop integrals in $n=4-2\epsilon$ dimensions:
\begin{eqnarray}
(4\pi)^2\int\frac{d^np}{(2\pi)^n}\frac{1}{[(p+q)^2+r][p^2+a]}%
&=&(4\pi)^{\epsilon}\left[\Gamma(\epsilon) -\int_0^1dx~\ln
g(a,r)\right]
\nonumber\\
(4\pi)^2\int\frac{d^np}{(2\pi)^n}\frac{p\cdot q}{[(p+q)^2+r][p^2+a]}%
&=&(4\pi)^{\epsilon}q^2\left[
-\frac{1}{2}\Gamma(\epsilon)+\int_0^1dx~x\ln g(a,r)\right]
\end{eqnarray}
where
\begin{eqnarray}
g(a,r)=q^2x(1-x)+rx+a(1-x)
\end{eqnarray}

Introducing the abbreviations,
\begin{eqnarray}
&&\bar g(a,r)=x(1-x)(1-y)+[rx+a(1-x)]y\nonumber\\
&&\tilde g_0=\frac{\bar g(0,r_4)}{\bar g(r_4,r_4)},~\tilde
g_1(r)=\frac{\bar g(1,r)}{\bar g(r_4,r)}\nonumber\\
&&\calG(r)=\frac{\ln\bar g(r_4,r)}{(r_4-1)r_4}%
+\frac{\ln\bar g(1,r)}{1-r_4}+\frac{\ln\bar g(0,r)}{r_4}
\end{eqnarray}
and denoting the parameter integrals in the form,
\begin{eqnarray}
X=\int_0^1dx\int_0^1dy~I[X]
\end{eqnarray}
where $X$ enumerates all of the defined integrals, the integrands
are
\begin{eqnarray}
I[\calX_2(r)]&=&\frac{x(1-y)\calG(r)}{y(1-y+r_4y)}\nonumber\\
I[\calX_1(r)]&=&\frac{1}{r_4-1}
\frac{x(1-y)}{y(1-y+r_4y)}\ln\tilde g_1(r)\nonumber\\
I[\calX_0]&=&\frac{1}{r_4-1}\frac{x(1-y)^2}{y^2(1-y+r_4y)} %
\ln\frac{\tilde g_1(0)}{\tilde g_1(r_4)}
\end{eqnarray}
for the $\calX$ sequence, and
\begin{eqnarray}
I[\calY_2(r)]&=&-\frac{\calG(r)}{1-y+r_4y}\nonumber\\
I[\calY_1(r)]&=&\frac{1}{1-r_4}\frac{1}{1-y+r_4y}\ln\tilde g_1(r)
\nonumber\\
I[\calY_0(r)]&=&\frac{1}{1-r_4}\frac{1-y}{y(1-y+r_4y)}\ln\tilde
g_1(r)\nonumber\\
I[\calU_0]&=&-\frac{1}{r_4}\frac{1-y}{y(1-y+r_4y)}\ln\tilde g_0
\nonumber\\
I[\calU_1]&=&-\frac{1}{r_4}\frac{1}{1-y+r_4y}\ln\tilde g_0
\end{eqnarray}
for the $\calY$ and $\calU$ sequences.

The above integrals have a magnitude of order one or smaller for
$r_4$ not very large, and can be readily integrated numerically.
This is a sufficient message for the first part of our numerical
analysis in section \ref{sec:numerical}. For the analysis in the
heavy mass limit, we need the leading terms of the integrals. We
obtain them in two ways. One is to use the techniques and formulae
developed already in the literature
\cite{vanderBij:1983bw,McDonald:2003zj}, and extend them slightly to
cover all cases occurring in our integrals. (There is a typographic
error in expansion (ii) on page 230 in Ref \cite{vanderBij:1983bw}:
$\frac{1}{2}\ln^2a$ should have a plus sign instead of minus.) The
leading terms can also be extracted directly. For illustration, we
calculate below the integrals $\calX_1(r_4)$ and $\calX_2(r_4)$ that
appear most frequently in eq. (\ref{eq:large1}). We finish first the
integration over $y$ in terms of logarithm and dilogarithm functions
using
\begin{eqnarray}
I(b)&=&\int_0^1\frac{dy}{y}\ln[1+(b-1)y]=-\Li_2(1-b)
\nonumber\\
J(b,r)&=&(r-1)\int_0^1dy\frac{\ln[1+(b-1)y]}{1+(r-1)y}
\nonumber\\
&=&\Li_2\left(\frac{b-r}{(b-1)r}\right)-\Li_2\left(\frac{b-r}{b-1}\right)
-\ln\frac{r-1}{b-1}\ln r+\frac{1}{2}\ln^2r
\end{eqnarray}
where $b>1,~r>1$. Denoting
\begin{eqnarray}
b_1=\frac{r_4x+1-x}{x(1-x)},~b_2=\frac{r_4}{x(1-x)},
~b_3=\frac{r_4}{1-x};~a_i=\frac{b_i-r_4}{b_i-1}
\end{eqnarray}
with $b_2\ge b_1\ge b_3\ge r_4>1>a_i>0$ for $x\in(0,1)$, and using
the abbreviations
\begin{eqnarray}
I_i=I(b_i),~J_i=J(b_i,r_4)
\end{eqnarray}
we express the integrals as follows:
\begin{eqnarray}
(r_4-1)\calX_1(r_4)&=&\int xdx\Big\{\left(\left[I_1-I_2\right]
-\left[J_1-J_2\right]\right)-\frac{J_1-J_2}{r_4-1}\Big\}
\nonumber\\
r_4(r_4-1)\calX_2(r_4)&=&\int xdx\Big\{
r_4\left(\left[I_2-I_1\right] -\left[J_2-J_1\right]\right)
-(r_4-1)\left(\left[I_2-I_3\right]-\left[J_2-J_3\right]\right)
\nonumber\\
&&-r_4\left[\frac{J_2-J_1}{r_4-1}-\frac{J_2-J_3}{r_4}\right]\Big\}
\end{eqnarray}
Since none of $I_i$ and $J_i$ diverges as a power as $r_4$ becomes
large, we have for $r_4\gg 1$,
\begin{eqnarray}
(r_4-1)\calX_1(r_4)&=&\int xdx\Big\{\left[I_1-I_2\right]
-\left[J_1-J_2\right]\Big\}+O(r_4^{-1})
\nonumber\\
r_4(r_4-1)\calX_2(r_4)&=&\int xdx\Big\{
r_4\left(\left[I_3-I_1\right] -\left[J_3-J_1\right]\right)
+\left[I_2-I_3\right]+\left[J_1-J_2\right]\Big\}+O(r_4^{-1})
\end{eqnarray}

To extract the leading terms, we have to expand the first
combination in $\calX_2(r_4)$ to $O(r_4^{-1})$ and all others to
$O(1)$. Consider the latter first. Since all $b_i\gg 1$ for $r_4\gg
1$, we use Landen identity of dilogarithm for the last two
combinations in $\calX_2$:
\begin{eqnarray}
I_2-I_3&=&\frac{1}{2}\ln(b_2b_3)\ln\frac{b_2}{b_3}
+\Li_2(1-b_2^{-1}) -\Li_2(1-b_3^{-1})
\nonumber\\
&=&-\frac{1}{2}\left[2\ln r_4-\ln x-2\ln(1-x)\right]\ln x
+O(r_4^{-1})
\nonumber\\
J_1-J_2&=&\left[\Li_2(a_1/r_4)-\Li_2(a_1)+\ln(b_1-1)\ln r_4\right]
-(a_1\to a_2;b_1\to b_2)
\nonumber\\
&=&\Li_2(1-x(1-x))-\Li_2(x)+\ln x\ln r_4+O(r_4^{-1})
\end{eqnarray}
Then
\begin{eqnarray}
B_2&\equiv&\int xdx\left\{[I_2-I_3]+[J_1-J_2]\right\}
\nonumber\\
&=&-\frac{1}{2}-\frac{11\pi^2}{36}+\frac{1}{12}\psi_1(1/6)
+\frac{1}{12}\psi_1(1/3)+O(r_4^{-1})
\nonumber\\
&\approx&0.435+O(r_4^{-1})
\end{eqnarray}
where $\displaystyle\psi_1(z)=\frac{d^2}{dz^2}\ln\Gamma(z)$ is the
trigamma function. Since
\begin{eqnarray}
\left[I_1-I_2\right]-\left[J_1-J_2\right]
=-[I_2-I_3]-[J_1-J_2]+O(r_4^{-1}),
\end{eqnarray}
this also gives the leading term
\begin{eqnarray}
r_4\calX_1(r_4)&=&-B_2+O(r_4^{-1})
\end{eqnarray}

The first combination in $\calX_2$ is more complicated. Using Landen
identity and expansions of $\Li_2(z)$ at $z=0$ and $z=1^-$, we have
\begin{eqnarray}
I_3-I_1&=&-\frac{1}{r_4}\frac{1-x}{x}\ln\frac{r_4}{1-x}+O(r_4^{-2})
\nonumber\\
J_3-J_1&=&\Li_2(a_1)-\Li_2(a_3)-\frac{1}{r_4}\frac{1-x}{x}\ln r_4
+O(r_4^{-2})
\end{eqnarray}
The combination is thus
\begin{eqnarray}
B_1&\equiv&r_4\int xdx\left\{\left[I_3-I_1\right]
-\left[J_3-J_1\right]\right\}
\nonumber\\
&=&\int dx\left\{(1-x)\ln(1-x)+r_4x\left[\Li_2(a_3)
-\Li_2(a_1)\right]\right\}+O(r_4^{-1})
\end{eqnarray}
The $\Li_2(a_i)$ terms can be worked out by integration by parts,
noting that $a_{1,3}=1$ at $x=1$ while $a_1=1$ and $a_3=0$ at $x=0$:
\begin{eqnarray}
\int xdx~\Li_2(a_{1,3})=\frac{\pi^2}{12}+\frac{1}{2}\int
dx~x^2\ln(1-a_{1,3})\frac{d\ln a_{1,3}}{dx}
\end{eqnarray}
where $\displaystyle\frac{d\Li_2(z)}{dz}=-\frac{\ln(1-z)}{z}$ is
applied. Upon expanding the integrand in $r_4^{-1}$, we arrive at
\begin{eqnarray}
\int xdx[\Li_2(a_3)-\Li_2(a_1)]&=&\frac{1}{2r_4}\int dx
\left[-\frac{\ln(1-x)}{x}(2x-2)+(1-x)\right]+O(r_4^{-2})
\nonumber\\
&=&\frac{5}{4r_4}-\frac{\pi^2}{6r_4}+O(r_4^{-2})
\end{eqnarray}
so that $\displaystyle B_1=1-\frac{\pi^2}{6}+O(r_4^{-1})$.

We collect below the leading terms for all integrals.
\begin{eqnarray}
\calX_0&=&\frac{\pi^2}{12}-\frac{1}{2}C_0+O(r_4^{-1})
\nonumber\\
r_4\calX_1(0)&=&\frac{1}{2}-\frac{\pi^2}{6}+O(r_4^{-1})
\nonumber\\
r_4\calX_1(r_4)&=&\frac{1}{2}+\frac{\pi^2}{12}-\frac{1}{2}C_0+O(r_4^{-1})
\nonumber\\
r_4^2\calX_2(0)&=&-1+\frac{1}{3}\pi^2-\ln r_4+O(r_4^{-1})
\nonumber\\
r_4^2\calX_2(r_4)&=&\frac{1}{2}-\frac{\pi^2}{4}+\frac{1}{2}C_0+O(r_4^{-1})
\end{eqnarray}
\begin{eqnarray}
r_4\calY_0(0)&=&\frac{\pi^2}{3}+O(r_4^{-1})
\nonumber\\
r_4\calY_0(r_4)&=&-\frac{\pi^2}{6}+C_0+O(r_4^{-1})
\nonumber\\
r_4^2\calY_1(0)&=& 1-\frac{\pi^2}{3}+\ln
r_4+\frac{1}{2}\ln^2r_4+O(r_4^{-1})
\nonumber\\
r_4^2\calY_1(r_4)&=&3-C_0+\ln r_4+O(r_4^{-1})
\nonumber\\
r_4^2\calY_2(0)&=&\frac{\pi^2}{3}+O(r_4^{-1})
\nonumber\\
r_4^3\calY_2(r_4)&=&-7+\frac{\pi^2}{2}+C_0 -2\ln
r_4+\ln^2r_4+O(r_4^{-1})
\end{eqnarray}
and
\begin{eqnarray}
r_4\calU_0&=&-\frac{\pi^2}{6}+C_0+O(r_4^{-1})
\nonumber\\
r_4^2\calU_1&=&3-C_0+\ln r_4+O(r_4^{-1})
\end{eqnarray}
with $C_0=2\sqrt{3}\textrm{Cl}(\pi/3)=-4\pi^2/9+1/6 \psi_1(1/6)+1/6
\psi_1(1/3)\approx 3.51586$, where $\textrm{Cl}$ is the Clausen
function. These leading terms have been numerically verified.

\vspace{0.5cm}
\noindent %

\end{document}